\documentclass[amssymb, amsmath, aps, prl, unsortedaddress, nofootinbib, twocolumn]{revtex4} 
\usepackage{bm,url,graphicx,natbib,fancyhdr}%

\long\def\symbolfootnote[#1]#2{\begingroup\def\thefootnote{\fnsymbol{footnote}}\footnote[#1]{#2}\endgroup}
\newcommand{\degr}{\ensuremath{^{\circ}}}
\newcommand{\ETHzs}{$E_{\rm{THz}}$\ }

\newcommand{\ETHzm}{E_{\rm{THz}}}

\newcommand{\OxPhysAddress}{University of Oxford, Department of Physics, Clarendon Laboratory, Parks
Road, Oxford, OX1 3PU, United Kingdom}
\newcommand{\ANUaddress}{Department of Electronic Materials Engineering, Research School of Physical Sciences and
Engineering, Institute of Advanced Studies, Australian National University, Canberra ACT
0200, Australia}

\begin{document}

\title{Influence of surface passivation on ultrafast carrier dynamics and terahertz radiation generation in GaAs}


\author{J.~Lloyd-Hughes} \email{james.lloyd-hughes@physics.ox.ac.uk}
\author{S.K.E.~Merchant} \address{\OxPhysAddress}
\author{L.~Fu}
\author{H.H.~Tan}
\author{C.~Jagadish} \address{\ANUaddress}
\author{E.~Castro-Camus}
\author{M.B.~Johnston} \email{m.johnston@physics.ox.ac.uk} \address{\OxPhysAddress}

\parbox{17.2cm}{\centering {PREPRINT: To appear in Appl. Phys. Lett. {\bf 89}, (Nov 2006)}}

\lhead[\thepage]{Lloyd-Hughes \emph{et al.}, To appear in Appl. Phys. Lett. {\bf 89},
(Nov 2006).}

\rhead[Lloyd-Hughes \emph{et al.}, To appear in Appl. Phys. Lett. {\bf 89}, (Nov
2006).]{\thepage}

\date{4th October 2006 (submission)}

\begin{abstract}{
The carrier dynamics of photoexcited electrons in the vicinity of the surface of
$\rm{(NH_4)_2S}$-passivated GaAs were studied via terahertz (THz) emission spectroscopy
and optical-pump THz-probe spectroscopy. THz emission spectroscopy measurements, coupled
with Monte Carlo simulations of THz emission, revealed that the surface electric field of
GaAs reverses after passivation. The conductivity of photoexcited electrons was
determined via optical-pump THz-probe spectroscopy, and was found to double after
passivation. These experiments demonstrate that passivation significantly reduces the
surface state density and surface recombination velocity of GaAs. Finally, we have
demonstrated that passivation leads to an enhancement in the power radiated by
photoconductive switch THz emitters, thereby showing the important influence of surface
chemistry on the performance of ultrafast THz photonic devices.}
\end{abstract}



\maketitle

Surface and interface states can dominate charge carrier transport in semiconductors, for
instance creating unexpectedly high mobilities in nanometre-thick silicon-on-insulator
structures\cite{Zhang06-286} or significant carrier trapping in polymer field-effect
transistors.\cite{pfet06} Marked improvements in the performance of macroscopic III-V
devices can be obtained by chemical treatments that remove the surface oxide layer and
passivate the semiconductor/air interface electrically and
chemically.\cite{SANDROFF87-277,YABLONOVITCH87-278} Typically, passivation prevents
electrons from surface atoms forming defect states within the semiconductor's
bandgap\cite{bessolov1998}, thereby reducing the surface recombination rate. Passivation
techniques have led to performance enhancements for III-V laser
diodes\cite{KAMIYAMA91-280}, solar cells\cite{MAUK89-279} and bipolar
transistors\cite{SANDROFF87-277}. However, discussion of passivation with regard to
sources of terahertz (THz) radiation has been limited to Schottky diode
multipliers\cite{Hartnagel02-281}, which produce continuous wave radiation at typically
$< 300$\,GHz. Surface states may also be expected to play an important role in broadband
emitters of THz radiation\cite{jqe24_255,apl83_3117}, where the photoexcited carrier
distribution lies within $\sim 1$\,$\mu$m of the surface.

In this paper we report an investigation into carrier recombination at
$\rm{(NH_4)_2S}$-passivated GaAs surfaces using time-resolved spectroscopy, and show how
this knowledge of carrier dynamics can be used to improve the performance of pulsed THz
emitters. Three complementary techniques were used in this study: (i) \emph{Surface THz
emission}, which is an excellent probe of the space-charge induced electric field at the
surface of bulk semiconductors, owing to a strong sensitivity to the bulk doping
level.\cite{prb02} (ii)\emph{Optical-pump THz-probe spectroscopy}, which allows the
conductivity of photocarriers in a semiconductor to be measured as a function of time
after photoexcitation.\cite{prl58_2355,jap90_5915} As the electron lifetime and mobility
can be determined using this technique, we are thus able to optimize materials for
specific device applications. (iii) \emph{THz emission from photoconductive switch
devices}, which we use as an example of how controlling the dynamics of charge carriers
in the vicinity of a surface, in particular by surface passivation, can be used to
improve significantly the performance of THz devices.

The surfaces of samples of semi-insulating (SI) GaAs and InSb [both (100) orientation,
with dark resistivities of 1.5$\times10^8$\,$\Omega$cm$^{-1}$ and
1.2$\times10^{-1}$\,$\Omega$cm$^{-1}$ respectively] were etched with 5:1:1 $\rm{H_2 SO_4
: H_2 O_2 : H_2 O}$, and subsequently passivated by dipping in $\rm{(NH_4)_2S}$ for
10\,minutes.\cite{YABLONOVITCH87-278} A reference set of samples were made from the same
wafers, without the passivation step, and were allowed to oxidize completely in air.


We used terahertz time-domain spectroscopy\cite{Schmuttenmaer04-98} to measure the THz
emission from passivated and etched samples of GaAs and InSb, in a setup similar to that
of Ref.\ \cite{prb04}. 90\% of the output of a Ti:Sapphire oscillator laser (10\,fs pulse
duration, 75\,MHz repetition rate, 450\,mW beam power, 790\,nm wavelength) was used to
generate carriers in the sample; the remainder was used to detect the emitted THz
transients using electro-optic sampling [with a 0.2\,mm $\left( 110\right)$ ZnTe crystal
on a 6\,mm $\left( 100\right)$ ZnTe substrate].

Terahertz emission from semiconductor surfaces can be used to investigate the carrier
dynamics in ion-damaged semiconductors\cite{prb04}, or to probe the surface charge
distribution. At the typical pump fluences available with unamplified Ti:sapphire lasers,
the dominant THz radiation mechanism in $\left( 100\right)$ GaAs is charge separation
under the surface field, while for higher mobility semiconductors such as $\left(
100\right)$ InAs it is the photo-Dember effect\cite{prb02} (the difference in electron
and hole mobility). At higher fluences, and for $\left( 110\right)$ and $\left(
111\right)$ crystal cuts, THz emission from optical rectification becomes
significant.\cite{Gu02-285}

\begin{figure}[bt]
    \centering
    \includegraphics[width=8.5cm]{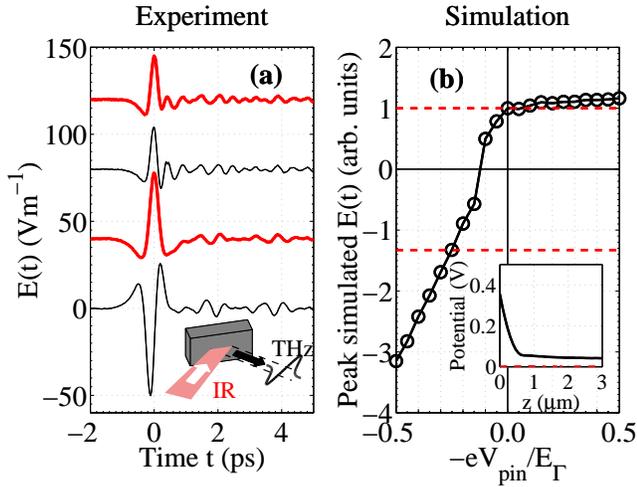}
    \caption{\label{FIG:surface_field}
(Color online) a) From bottom to top: Emitted time-domain THz electric field from
surfaces of etched GaAs, passivated GaAs, etched InSb and passivated InSb -- the etched
samples act as references. The peak of THz pulses emitted from InAs (not shown) was
$+550$\,Vm$^{-1}$. The oscillations after the main pulse result from the THz absorption
lines of atmospheric water vapour. Passivation produces no noticeable change in \ETHzs
for InSb, but causes the polarity to flip and the amplitude to decrease for GaAs. Inset:
schematic of experimental geometry, showing the infra-red (IR) emitter pump beam at
45\degr\ to the emitter, and the radiated THz pulse. b) Peak of simulated THz electric
field emission from GaAs surfaces, versus surface pinning potential energy $eV_{\rm pin}$
relative to the bandgap energy $E_{\Gamma}=1.42$\,eV. Dotted horizontal lines represent
the relative measured peak fields for the passivated (top) and etched reference (bottom)
samples. Inset: Simulated electrostatic potential 50\,fs after the arrival of the
infrared pump pulse, as a function of depth $z$ into the semiconductor, for $V_{\rm
pin}=0.355$\,V (solid line) and $V_{\rm pin}=0$\,V (dashed line).
    }
\end{figure}

The THz emission from the surfaces of the passivated and reference samples was measured,
and is shown in Fig.~\ref{FIG:surface_field}(a). The polarity of the THz electric field
from the etched GaAs sample was opposite to that of InAs (not shown), while for
passivated GaAs the radiated pulses had the same polarity as InAs. The polarity change
suggests that passivation suppresses the surface states that create the surface field,
namely that passivated GaAs acts as a photo-Dember emitter. No significant change in the
THz emission from samples of InSb was observed after applying the same passivation
process, since InSb (like InAs) is primarily a photo-Dember emitter
(Fig.~\ref{FIG:surface_field}).\cite{Gu02-285}

We have used a three-dimensional carrier dynamics simulation\cite{prb02} to investigate
how changes to the surface states in GaAs alter THz emission. The influence of surface
defects can be described by the pinning of the electrostatic potential at the surface,
where the potential relative to the bulk is $V_{\rm pin}$.
Fig.~\ref{FIG:surface_field}(b) indicates the peak of the simulated THz electric field as
$V_{\rm pin}$ is varied. With no Fermi level pinning ($V_{\rm pin}=0$) the simulated THz
radiation has the same sign as InAs, and the semiconductor acts as a photo-Dember emitter
-- there is no surface field, as the inset to Fig.~\ref{FIG:surface_field}(b) indicates.
As $V_{\rm pin}$ becomes increasingly negative the simulated field strength changes in
sign, owing to the surface field component. Therefore, assuming that the passivated GaAs
sample has $V_{\rm pin}=0$, the pinning potential in the etched sample can be estimated
from the relative emission amplitudes as $V_{\rm pin}=-0.25 E_{\Gamma}/e=0.355$\,V.

In order to investigate the dynamics of photo-excited carriers close to surface defects
we measured the time-resolved conductivity $\sigma(t')$ of the passivated and etched GaAs
samples. The experimental geometry used was as follows: 45\% of the laser's output was
used to generate THz pulses from a SI-GaAs photoconductive
switch\cite{Schmuttenmaer04-98}, and 10\% to detect the transient after transmission
through the sample. The remaining 45\% of the beam was used to photoexcite the sample
colinearly -- this sample pump beam was mechanically chopped at 160\,Hz. The change in
the transmitted THz electric field induced by the pump was recorded as a function of the
arrival time $t'$ of the sample pump pulse relative to the THz pulse.

\begin{figure}[tb]
    \centering
    \includegraphics[width=8.5cm]{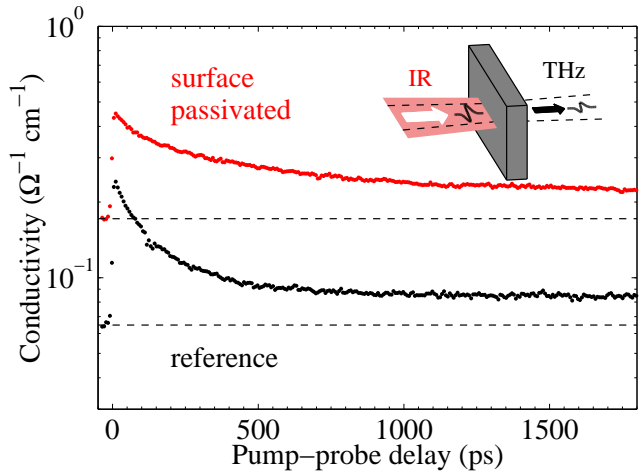}
    \caption{\label{FIG:conductivity}
    (Color online) Time-resolved conductivity of passivated (top) and reference (bottom) GaAs
    samples, as measured via optical-pump, THz-probe spectroscopy. The dotted lines
    indicate a non-zero conductivity before the pump pulse arrives, owing to the bulk
    lifetime ($\tau_b=15$\,ns) exceeding the repetition period between laser pulses
    (13.3\,ns). Inset: schematic of experimental geometry, showing the infra-red (IR) sample pump beam,
    and the incident and transmitted THz probe pulse.
    }
\end{figure}

The time-resolved conductivity $\sigma(t')$ was readily obtained from this
data\cite{Beard00-228}, and is shown in Fig.~\ref{FIG:conductivity}. At zero pump-probe
delay time ($t'=0$) the conductivity increases rapidly owing to the photogeneration of
electrons. The decay in conductivity is non-exponential: at early delay times surface
recombination significantly depopulates the electron concentration, while at later delay
times ($\gtrsim 600$\,ps) the carrier distribution has had time to diffuse into the bulk,
reducing the role of surface recombination.\cite{Beard00-228} It can be seen that the
surface passivated sample has a larger initial conductivity than that of the etched
sample, and a longer initial decay time constant. As the incident photon flux was
identical for the two samples, this increase in conductivity can be attributed to a $1.9
\times$ larger initial electron mobility $\mu$. An exponential fit to the initial decay
(up to 400\,ps) produces a time constant of $\tau=389$\,ps for the passivated sample,
twice that of the etched sample ($\tau=192$\,ps). We observed a comparable enhancement in
conductivity using Na$_2$S$\cdot$9H$_2$O to passivate the surface of
GaAs.\cite{YABLONOVITCH87-278}

Sulfur passivated GaAs is known to be partially unstable in oxygen -- indeed after two
days the peak conductivity had decayed by 9\,\%. The deposition of a thin layer of
silicon nitride after sulfur passivation may prevent the degradation of the
sulfur-treated GaAs surface.\cite{shikata1991,hobson1993} The effect of this degradation
was minimized during these experiments by storing samples in a nitrogen glove box.

We modelled the non-exponential shape of the decay in $\sigma$ using a solution to the 1D
diffusion equation\cite{Beard00-228} in order to obtain the surface recombination
velocity $S_0$. With a bulk lifetime $\tau_b=15$\,ns (taken from the limit of the decay
in Fig.~\ref{FIG:conductivity}) good agreement is found with the measured $\sigma$ when
$S_0=1.2\times 10^6$\,cm\,s$^{-1}$ for the etched reference, and $S_0=2.0\times
10^5$\,cm\,s$^{-1}$ for the surface passivated sample. These values correspond well to
those in the literature for etched and passivated GaAs surfaces (however $S_0$ can be
reduced further to $S_0=10^3$\,cm\,s$^{-1}$ by alternative surface
treatments\cite{YABLONOVITCH87-278}). The Shockley-Read-Hall model predicts that $S_0=n_t
v \Sigma$ for a surface areal trap density $n_t$, scattering cross-section $\Sigma$ and
carrier velocity $v$. Assuming that $v$ and $\Sigma$ are identical before and after
passivation, $n_t$ for the passivated sample is 17\% of that in the etched sample. The
passivation step can therefore directly be seen to produce a surface with fewer
recombination centres.

At large pump-probe delay times ($\gtrsim 1000$\,ps) the decay in conductivity slows, as
the carrier distribution has diffused into the bulk. The bulk lifetime exceeds the
repetition period between laser pulses (13.3\,ns), resulting in a non-zero conductivity
at negative pump-probe delays.

\begin{figure}[tb]
    \centering
    \includegraphics[width=8.5cm]{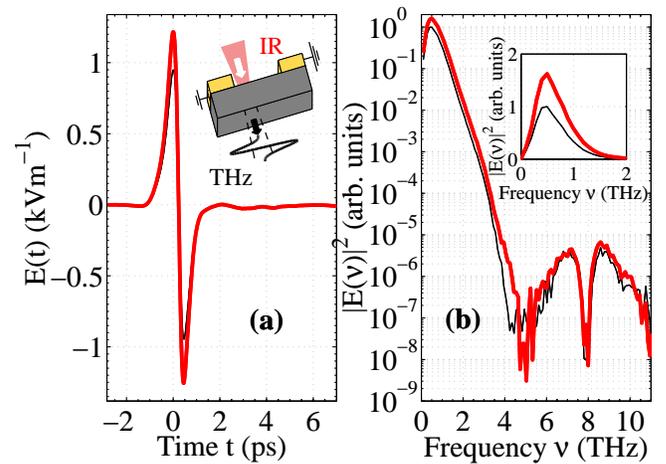}
    \caption{\label{FIG:photoconductive_switch}
(Color online) (a) Electric field strength of emitted THz pulses from 400\,$\mu$m-gap
photoconductive switches made on passivated GaAs (thick line) and an etched reference
(thin line), as a function of electro-optic delay time. Inset: schematic of experimental
geometry, showing the infra-red (IR) emitter pump beam close to the anode contact of the
photoconductive switch, and the radiated THz pulse. (b) Power spectra of THz emission
from passivated (thick line) and etched (thin line) GaAs, obtained by Fourier
transforming the data in (a). These data are shown on a linear scale in the inset.}
\end{figure}

The THz emission from large-area photoconductive switches fabricated on passivated and
etched GaAs is reported in Fig.~\ref{FIG:photoconductive_switch}. It can be seen that the
peak THz electric field strength from the passivated sample is larger than that of the
etched reference, with a near doubling of the emitted power
[Fig.~\ref{FIG:photoconductive_switch}(b)]. This increase is due to the larger change in
conductivity $\sigma$ of the passivated sample (owing to its greater mobility), since the
emitted THz electric field is $\ETHzm \propto \partial J/\partial t=\partial(\sigma
E)/\partial t$, where current density $J$ flows at an applied field $E$.\cite{prb02}

The observation that electrons in SI-GaAs can have lifetimes exceeding the pulse period
of high repetition rate lasers (Fig.~\ref{FIG:conductivity}) is of significance to
photoconductive THz emitters and detectors. The residual carriers created by the
preceding laser pulse will reduce the change in conductivity, and therefore also the
emitted field strength. In addition, the lowering of the dark resistivity increases the
noise background and heats the emitter, which can require water-cooling when operated at
high voltages.\cite{rsi73_1715} Similarly, the noise in photoconductive detectors of THz
radiation is increased by long-lived electrons in devices fabricated on SI-GaAs, and
low-temperature grown or ion-damaged layers (thinner than the absorption depth) on
semi-insulating substrates.\cite{Hussain06-240}

In conclusion, we have investigated the ultrafast carrier dynamics of passivated GaAs
surfaces via time-resolved conductivity measurements, THz emission spectroscopy, and
simulation. After passivation the THz electric field emitted from the GaAs surface
flipped in polarity to correspond to that of photo-Dember emitters such as InSb and InAs.
This change is indicative of the removal of the surface defects after passivation, and
was reproduced by carrier dynamics simulations of terahertz emission. Additionally, the
mean mobility of photoexcited electrons in $\rm{(NH_4)_2S}$-passivated GaAs was measured
by optical-pump terahertz-probe spectroscopy, and was found to be twice that of an
unpassivated reference sample. Ensuring a high-quality surface with a low defect
concentration was shown to enable improved photoconductive sources of THz radiation, as
demonstrated by the observation of a power enhancement for photoconductive antenna
emitters after passivation. This method can be used in addition to other schemes that
increase the power of THz sources (such as placing a hemispherical silicon lens to
collimate the emitted radiation, or putting an anti-reflection coating to enhance
coupling from the emitter into free space), and has the benefit of introducing no
dispersive media into the THz path.

The authors would like to acknowledge support from the EPSRC (UK), the Royal Society (UK)
and the ARC (Australia) for this work.



\end{document}